\documentclass[conference]{IEEEtran}
\IEEEoverridecommandlockouts
\usepackage[utf8]{inputenc}
\usepackage[T1]{fontenc}
\usepackage{comment}
\usepackage{cite}
\usepackage{amsmath,amssymb,amsfonts}
\usepackage{algorithmic}
\usepackage[linesnumbered,ruled]{algorithm2e}
\usepackage{graphicx}
\usepackage{textcomp}
\usepackage{xcolor}
\usepackage{booktabs}
\usepackage{listings}
\usepackage{mdframed}
\usepackage{cleveref}
\usepackage{breqn}
\usepackage{stfloats}
\usepackage{afterpage}
\usepackage{float}
\usepackage{here}
\usepackage{caption}
\usepackage{subcaption}
\usepackage{url} 
\usepackage{enumitem}

\usepackage[most]{tcolorbox} % 枠付きのボックスを作成するパッケージ
\def\BibTeX{{\rm B\kern-.05em{\sc i\kern-.025em b}\kern-.08em
    T\kern-.1667em\lower.7ex\hbox{E}\kern-.125emX}}

\usepackage{tikz}
\usepackage{lipsum} % ←本文デモ用、不要なら削除
\makeatletter
\def\ieeecopyright{
  \footnotesize
  © 2025 IEEE. Personal use of this material is permitted.}
\makeatother
\AddToHook{shipout/firstpage}{%
  \begin{tikzpicture}[remember picture,overlay]
    \node[anchor=south west,xshift=1.0cm,yshift=0.8cm]
      at (current page.south west)%
      {\parbox{\linewidth}{\raggedright\ieeecopyright}};
  \end{tikzpicture}%
}
\crefname{figure}{Fig.}{Figures}
\crefname{table}{TABLE}{Tables}
\crefname{section}{Section}{Sections}
\crefname{equation}{eq.}{eqs.}

\begin{document}

\newcommand{\TOOLNAME}{\textit{Function-as-Subtask}}
\newcommand{\SHORTNAME}{\textit{FasS}}

\title{
Work-in-Progress: Function-as-Subtask API Replacing Publish/Subscribe for OS-Native DAG Scheduling
}

\author{
  Takahiro Ishikawa-Aso$^{\dagger\ddagger}$,
  Atsushi Yano$^{*\ddagger}$,
  Yutaro Kobayashi$^{\ddagger}$,
  Takumi Jin$^{\ddagger}$,
  Yuuki Takano$^{\ddagger}$,
  Shinpei Kato$^{\dagger}$
  \\
  $^{\dagger}$The University of Tokyo, Japan \space
  $^{*}$Saitama University, Japan \space
  $^{\ddagger}$TIER IV Incorporated, Japan
}

\lstdefinestyle{mystyle}{
    basicstyle=\ttfamily\footnotesize,
    breakatwhitespace=false,
    breaklines=true,
    captionpos=b,
    keepspaces=true,
    numbers=left,
    numbersep=10pt, % 行番号とコードの間に空白を設ける
    showspaces=false,
    showstringspaces=false,
    showtabs=false,
    tabsize=2,
    framexleftmargin=15pt, % 行番号がボックスからはみ出さないようにする
    xleftmargin=15pt,
    morekeywords={f0, f1, f2, topic0, topic1, topic2, MulRecv, MulSend, VecToPubs, VecToSubs}, % ハイライト
    keywordstyle=\color{red}\bfseries, % キーワードの色を赤にし、太字に設定
}

\lstset{style=mystyle}

\newtcolorbox{codebox}[1][]{
    enhanced,
    unbreakable,
    colback=white,
    colframe=black,
    boxrule=0.5pt,
    sharp corners,
    boxsep=2pt,
    left=5pt,
    right=5pt,
    top=2pt,
    bottom=2pt,
    listing only,
    listing options={style=mystyle},
    overlay={\node[anchor=south east, outer sep=2pt, font=\footnotesize] at ([xshift=-3pt, yshift=3pt]frame.south east) {#1};}
}

\maketitle

\begin{abstract}
The Directed Acyclic Graph (DAG) task model for real-time scheduling finds its primary practical target in Robot Operating System~2 (ROS~2). However, ROS~2’s publish/subscribe API leaves DAG precedence constraints unenforced: a callback may publish mid-execution, and multi-input callbacks let developers choose topic-matching policies. Thus preserving DAG semantics relies on conventions; once violated, the model collapses. We propose the Function-as-Subtask (FasS) API, which expresses each subtask as a function whose arguments/return values are the subtask’s incoming/outgoing edges. By minimizing description freedom, DAG semantics is guaranteed at the API rather than by programmer discipline. We implement a DAG-native scheduler using FasS on a Rust-based experimental kernel and evaluate its semantic fidelity, and we outline design guidelines for applying FasS to Linux sched\_ext.
\end{abstract}

\vspace{1mm}


\vspace{-2.5mm}

\section{Introduction}
\begin{figure}[tb]
  \centering
  \includegraphics[width=\linewidth]{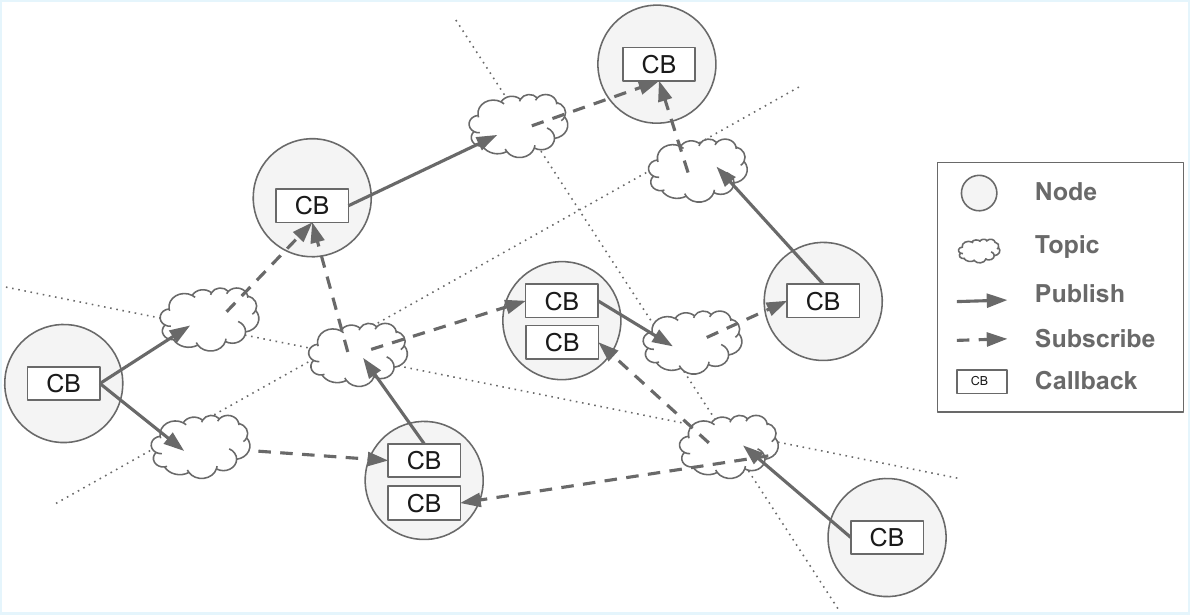}
  \vspace{-7mm}
  \caption{Component-oriented real-time system (e.g., ROS~2).}
  \label{fig-component-oriented-realtime}
  \vspace{-6mm}
\end{figure}

\vspace{-1.5mm}
Many modern autonomous cyber‐physical systems (CPS) are modeled as component‐oriented real‐time systems (\cref{fig-component-oriented-realtime}), mostly deployed on Robot Operating System~2 (ROS~2)~\cite{ros2_repo}.
A typical system consists of independent nodes that publish to or subscribe to topics defined by data types and identifiers.
Each node’s callbacks receive messages from subscribed topics and publish messages to others, collectively forming integrated dataflows throughout the system.
In order to guarantee the safety of such systems, timing analysis is required.

The Directed Acyclic Graph (DAG) task model provides a theoretical basis for timing analysis of component‐oriented real‐time systems.
DAG-based real-time scheduling was already established before ROS and AUTOSAR emerged\cite{kwok1999static, baruah2012generalized, verucchi2023survey}.
With their advent, such middleware became targets for DAG-scheduling theory, with ROS~2 now being representative \cite{casini2019response, choi2020autosar, teper2022end}.
Rather than DAG models, most scheduling analyses within ROS~2 follow cause--effect chains---trigger sequences formed by topics and timers---under Executors (middleware-level schedulers).
Meanwhile, emerging efforts aim to apply traditional DAG-scheduling model to ROS~2 as it is~\cite{yano2024work, ishikawa2025work}.

However, the publish/subscribe API leaves precedence constraints within a DAG unenforced, compelling OSes to rely on programmer's adherence to conventions for OS-native DAG scheduling.
As the API leaves \texttt{publish()} placement unconstrained, messages may be emitted mid-callback, violating the DAG semantics of ``completion $\!\rightarrow\!$ successor release''.
Moreover, when a callback takes multiple inputs, the application must specify a matching policy (\textit{message\_filters}\cite{ros_message_filters_wiki}) for messages from multiple topics, where embedded timestamps further affect matching logic.
Thus, preserving DAG semantics---referred to in this paper as \textbf{completion boundaries and join synchronization}---relies on development conventions; once violated or ignored, the DAG task model collapses.

We propose \textit{\textbf{Function-as-Subtask (FasS) API}}, an interface for native DAG-task-model support.
The \SHORTNAME{} API expresses each subtask in DAG tasks as a simple function, whose arguments and return values correspond to its incoming and outgoing edges.
By granting programmers minimal freedom, OSes can ensure DAG-model semantics without relying on their discipline.
Similar issues were reported in OpenMP: its flexible \textit{tied} and \textit{depend} semantics made model‐consistent scheduling unreliable, leading to constrained scheduling policies~\cite{sun2017real, sun2020real}.
Whereas ROS once adapted DAG-scheduling concepts into its own model, we offer a ROS-like middleware guideline enabling direct use of pure DAG-scheduling theory.
We implement a DAG-native scheduler using the \SHORTNAME{} API on our Rust-based experimental kernel and propose design guidelines for applying it to Linux’s sched\_ext framework~\cite{linux-sched-ext-doc}.

This paper makes the following contributions:
\vspace{-1mm}
\begin{enumerate}
\item
We propose \TOOLNAME{} (\SHORTNAME{}) API, enabling function-based descriptions of subtasks in DAG task models for task schedulers to natively support such models.
\item
\SHORTNAME{}-based systems free system software from depending on developer discipline, unlike publish/subscribe systems that had to trust it to treat applications as DAG tasks.
\item
We present case studies showing how the \SHORTNAME{} API facilitates native DAG-task-model support in our Rust ``Future-as-Thread'' OS and Linux sched\_ext.
\end{enumerate}

\section{Model GAP between ROS~2 Publish/Subscribe and DAG Task} \label{section-model-gap}

\subsection{DAG Task Semantics for Real-time Scheduling}
\label{subsection-dagsched-model}

Let an application be a DAG $G=(V,E)$ whose vertices $v\in V$ are \emph{subtasks} and edges $(u,v)\in E$ encode precedence constraints. For the $k$-th job, each subtask instance has an activation (ready) time $s_v^k$ and a finish time $f_v^k$.

\textbf{Completion boundary.}
The completion of a subtask is the instant $f_v^k$ at which its execution ends; no effect of $v$ may be released before $f_v^k$. Formally, successors cannot start before the predecessor has completed:
\vspace{-1mm}
\[
  \forall (u,v)\in E:\quad s_v^k \;\ge\; f_u^k .
\]
\vspace{-5mm}

\textbf{Join synchronization (multi-predecessor readiness).}
If $v$ has multiple predecessors, its readiness is gated by the \emph{AND} of their completions for the same job instance:
\vspace{-1mm}
\[
  \forall v\in V:\quad
  s_v^k \;\ge\; \max_{(u,v)\in E} f_u^k .
  \vspace{-1.5mm}
\]
Thus, $v$ becomes eligible to execute exactly when \emph{all} predecessors have completed; no partial or timestamp-based merge.

\textbf{Job completion.}
Let $S\subseteq V$ be the sinks (no successors). The job completion time is
\vspace{-2mm}
\[
  f_G^k \;\triangleq\; \max_{v\in S} f_v^k ,
  \vspace{-1.5mm}
\]
which is equivalently $\max_{v\in V} f_v^k$. The job meets its deadline iff $f_G^k \le r_k + D$ ($r_k$: release time; $D$: relative deadline).

These rules capture the two semantics used later to assess API suitability: (i) effects are released only at subtask \emph{completion} (tail of the subtask), and (ii) \emph{joins} are strict precedence-based synchronizations (logical AND over predecessors), not policy-dependent matches; in the ideal model, the successor becomes ready \emph{immediately} upon the last predecessor’s completion (i.e., zero join latency, \( s_v^k = \max_{(u,v)\in E} f_u^k \)).

\subsection{Limitations of Publish/Subscribe API for DAG Task Model} \label{subsection-pubsub-model}
\vspace{-1mm}
ROS~2 systems are component-oriented real-time systems (\cref{fig-component-oriented-realtime}) in which the realized DAGs are determined by the publish/subscribe API exposed to application programmers.
Each node comprises callbacks; a callback executes upon receiving messages on its subscribed topics and then publishes to downstream topics.
Through this relay of publish/subscribe interactions, system-wide dataflows (DAGs) are formed.

\cref{subsection-dagsched-model} formalized semantics for assessing API suitability as a model interface for real-time DAG scheduling: completion boundary and join synchronization.
The ROS~2 publish/subscribe API permits code that can violate these semantics; consequently, an OS aiming for DAG-native scheduling must rely on programmers’ discipline.

The first semantics (completion boundary) is vulnerable because the ROS~2 API does not constrain where \texttt{publish()} may be invoked inside a callback.
If a message is published mid-callback, a successor can be released before the subtask has completed, contradicting the completion boundary.
Even when all \texttt{publish()} calls are placed at the tail of the callback, multiple publications can still violate the intended semantics: depending on QoS settings, an earlier \texttt{publish()} may block, delaying subsequent \texttt{publish()} calls.
Consequently, not all outgoing edges are released at completion, effectively introducing post-completion delay.

The second semantics (join synchronization) is also fragile because multi-input callbacks in ROS~2 must select a matching policy via \textit{message_filters}\cite{ros_message_filters_wiki}.
Both \emph{ExactTime} (triggering only on identical timestamps) and \emph{ApproximateTime} (admitting adaptive tolerance) match on message timestamps; thus, join readiness depends on developer-chosen policy and timestamp management.
Additionally, zero-copy IPC middleware should be used to achieve zero join latency\cite{ishikawa2025ros}.

\vspace{-2mm}
\section{Function-as-Subtask API} \label{section-fas}
\vspace{-1mm}
In this section, we define the \TOOLNAME{} (\SHORTNAME{}) API and illustrate it through two OS examples.
% ラベルは左端に固定 / テキスト開始位置は labelsep で調整
\begin{enumerate}[label=Case\arabic*),
                  widest=Case9),
                  labelindent=0pt,
                  leftmargin=!,      % = labelindent + labelwidth + labelsep
                  align=left,
                  labelsep=0.7em]    % ← ここで“普通の enumerate”っぽさに寄せる
\item
\textbf{Our Rust-based experimental kernel\footnote{\url{ https://github.com/tier4/awkernel.}}}:
A single-address-space OS leveraging Rust ownership for isolation \cite{narayanan2020redleaf}.
Threads are implemented as futures, and the scheduler advances execution by polling them.
\item \textbf{Linux kernel with sched\_ext}: sched\_ext is a scheduling class merged into the mainline in version 6.12, which enables custom schedulers via eBPF programs.
\end{enumerate}

\vspace{-3mm}
\subsection{Specification of the FasS API} \label{subsection-fas-1}
\vspace{-1mm}
The \SHORTNAME{} API is a programming model that represents each subtask in the DAG task model as a simple function.
Intuitively, a function’s arguments and return values correspond to the subtask’s incoming and outgoing edges, respectively (\cref{fig-fas-simple-example}).
Strictly speaking, they correspond to topics—data channels—so a single argument may receive data from multiple incoming edges, and a single return value may feed multiple outgoing edges.
The \SHORTNAME{} API is defined as an interface through which an application programmer provides the following three pieces of information:
\vspace{-3mm}
\begin{figure}[tbh]
  \centering
  \includegraphics[width=1.0\linewidth]{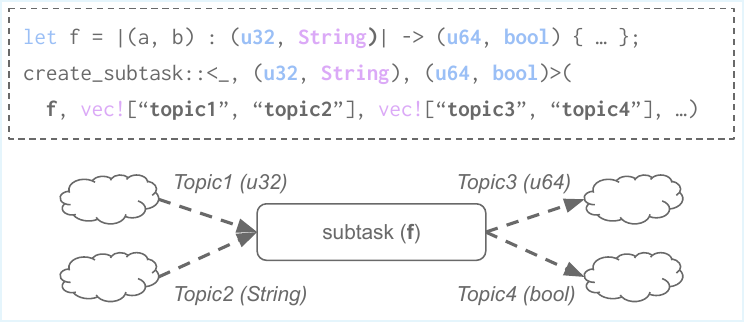}
  \vspace{-5mm}
  \caption{Basic example of Function-as-Subtask API.}
  \label{fig-fas-simple-example}
  \vspace{-2mm}
\end{figure}
\begin{enumerate}
\item
\textbf{A function representing the subtask}: Its arguments correspond to incoming edges and its return values to outgoing edges.
Within the function body, no operations that modify or expose the DAG structure—such as explicit \texttt{publish()} calls—are allowed.
\item 
\textbf{Lists of topics}: Corresponding to the function’s arguments (inputs) and return values (outputs).
\item
\textbf{Attributes}: Additional parameters required by the DAG scheduling algorithm.
In a static-priority–based DAG scheduling, each \SHORTNAME{} is assigned a priority value here.
\end{enumerate}

By limiting the API to the minimal expressiveness for subtask construction, the scheduler need not trust programmers to follow conventions or understand system internals.
Even without such knowledge, their code can be safely treated as a DAG task model.
Compared with the publish/subscribe API, DAG construction and control are delegated to a lower layer, removing excessive freedom.
Although application code may include elements that affect scheduling behavior—such as synchronized accesses to shared variables or blocking system calls—their handling is left for future work (\cref{section-conclusion}).

In the \SHORTNAME{}-based library design, the DAG construction phase should be clearly separated from the post-commit scheduling phase.
At commit time, the system verifies that the constructed DAG satisfies schedulable structures and attributes.
For example, if multiple subtasks publish messages to the same topic, independent DAGs are formed for each source, which may be considered invalid depending on the DAG construction policy.
\Cref{fig-faas-lib-example} shows an API example in the Case1 OS where the methods creating source, intermediate, and sink subtasks are clearly separated, reflecting the design principle guiding users to form a one-source–one-sink DAG.

\vspace{-2mm}
\begin{figure}[tbh]
\begin{codebox}[Rust]
\begin{lstlisting}[]
let dag = create_dag();
dag.register_periodic_subtask::<_, (i32,)>(
    f0, // || -> (i32,)
    vec!["topic0"], period, ...);

dag.register_subtask::<_, (i32,), (i32, i32)>(
    f1, // |(i32,)| -> (i32, i32)
    vec!["topic0"],
    vec!["topic1", "topic2"], ...);

dag.register_sink_subtask::<_, (i32, i32)>(
    f2, // |(i32, i32)| -> ()
    vec!["topic1", "topic2"], ...);

finish_create_dags(&[dag.clone()]);
\end{lstlisting}
\end{codebox}
\vspace{-2mm}
\caption{Function-as-Subtask API app. code example.}
\label{fig-faas-lib-example}
\end{figure}

\vspace{-4mm}
\subsection{Justification of the FasS API’s Sufficiency} \label{subsection-fas-2}
\vspace{-1mm}
Implementing a scheduler that natively supports the DAG task model under the \SHORTNAME{} API is feasible in languages with type-directed reification and static dispatch.
In other words, the system must automatically generate subscription and publisher objects in the publish/subscribe model from the type information of a \SHORTNAME{} function’s arguments and return values.
\Cref{fig-awkernel-code} shows the subtask-creation function implemented in our Rust-based experimental kernel (Case1).
Using Rust macros and templates, code for subscriptions and publishers is generated at compile time for any combination of argument and return types.
The same approach applies to other environments, such as Linux (Case2) or languages like C++.

\begin{figure}[tb]
\begin{codebox}[Rust]
\begin{lstlisting}[]
async fn create_subtask<F, Args, Ret>(
  f: F, ...,
) where
  F: Fn(<Args::Subs as MulRecv>::Items)
    -> <Ret::Pubs as MulSend>::Items
    + Send + 'static,
  Args: VecToSubs, Ret: VecToPubs,
  Args::Subs: Send, Ret::Pubs: Send,
{
 let subs = <Args as VecToSubs>::gen_subs(..);
 let pubs = <Ret as VecToPubs>::gen_pubs(..);
 loop {
   let args: <<Args as VecToSubs>::Subs as
     MulRecv>::Items = subs.recv_all().await;
   let results = f(args);
   pubs.send_all(results).await;
 }
}

trait MulRecv {
 type Items;
 fn recv_all(&self) -> Pin<Box<dyn Future<Output=Self::Items> + Send + '_>>;
}
trait MulSend {
 type Items;
 fn send_all(&self, items: Self::Items) -> Pin<Box<dyn Future<Output=()> + Send + '_>>;
}
trait VecToSubs {
 type Subs: MulRecv;
 fn gen_subs(...) -> Self::Subs;
}
trait VecToPubs {
 type Pubs: MulSend;
 fn gen_pubs(...) -> Self::Pubs;
}
// Macros and templates implement recv_all, send_all, gen_subs, and gen_pubs for all Function-as-Subtask arg/ret types.
...
\end{lstlisting}
\end{codebox}
\vspace{-1mm}
\caption{Function-as-Subtask API app. code example2.}
\label{fig-awkernel-code}
\vspace{-6mm}
\end{figure}

\vspace{-1mm}
\subsection{Reference OS Designs Based on the FasS API} \label{subsection-fas-3}
\begin{figure}[tb]
  \centering
  \includegraphics[width=\linewidth]{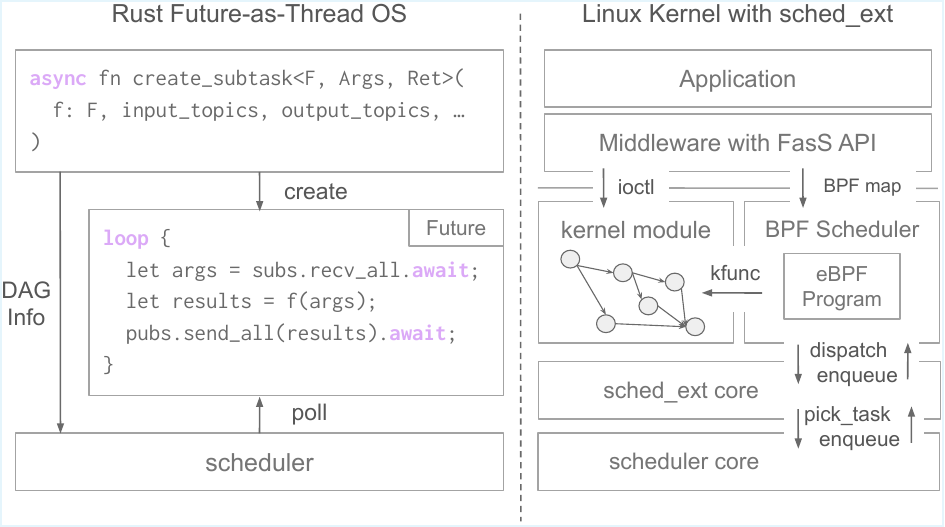}
  \vspace{-8mm}
  \caption{Reference OS designs based on the FasS API.}
  \label{fig-os-design}
  \vspace{-6mm}
\end{figure}

\vspace{-1mm}
We implement a \emph{FasS}-based DAG scheduler in our Rust single-address-space experimental OS (\cref{fig-os-design} left).
Threads are realized as \emph{Futures}; with no user–kernel boundary, subtasks created via \emph{FasS} (Fig.~4) are passed directly to the scheduler.
From the \emph{FasS} API information, the kernel derives the DAG structure and precedence constraints and enforces \emph{completion boundary} and \emph{join synchronization}.
Thus, scheduling natively supports the DAG task model without relying on programmers to follow conventions or understand system internals.

Fig.~5 (right) outlines an ongoing design of a DAG scheduler on Linux \textit{sched\_ext}~\cite{linux-sched-ext-doc}.
\textit{sched\_ext} exposes an eBPF-based scheduling class (\textit{ext\_sched\_class}) whose callbacks are implemented through \texttt{sched\_ext\_ops} powered by \texttt{struct\_ops}.
Implementing the DAG algorithm fully in eBPF is limited by the verifier, so we plan to keep eBPF as a thin entry and delegate the core scheduler to kernel-module functions exported as \emph{kfunc}s.
The DAG graph and parameters are registered from userspace (e.g., via \texttt{ioctl}).
Whereas Case1 maps one \emph{FasS} to one \emph{Future}, Case2 assigns a dedicated thread per \emph{FasS} and realizes an analogous control loop under \textit{sched\_ext}.
In both cases, the design rests on the same premise: \emph{FasS} lets the OS satisfy DAG semantics by API constraint rather than developer discipline.

\vspace{-2mm}
\section{Evaluation} \label{section-evalation}
\vspace{-2mm}
We assess join synchronization (\cref{subsection-dagsched-model}) to quantify \emph{FasS} API's benefit (DAG semantics guaranteed at the API, not by programmer discipline).
The metric is \emph{Join Latency}, defined here as the time from all index-\(i\) messages being published to the \(i\)-th join callback becoming \emph{ready}.

\textbf{Baselines/conditions:}
(1) \emph{Case1 kernel} with DAG scheduling via \SHORTNAME{} API;
(2) \emph{ROS~2} with \emph{ExactTime} or \emph{ApproximateTime} with \(\emph{Max Interval Duration}\in\{10,30,50\}\,\mathrm{ms}\) (maximum allowed timestamp difference for matching).
Semantics-compliant conditions are \emph{Case1 kernel} and \emph{ROS~2+ExactTime} (with correct stamping); \emph{ROS~2+ApproximateTime} is a stress case that emulates misuse within API freedom.

\textbf{Workload:}
single-level fork--join DAG (one source\(\rightarrow\)\(N\) parallel subtasks\(\rightarrow\)one sink), period \(25\,\mathrm{ms}\); each parallel subtask executes uniformly in \([1,50]\,\mathrm{ms}\); \(N\in\{2,4,6\}\).

\textbf{ROS~2 setup:}
single process MultiThreadedExecutor to remove IPC; highest priority for the application.
Timestamp policy: the source sets \texttt{header.stamp}; an intermediate subtask updates the stamp only under \emph{ApproximateTime} (to the publish time) and does not update it under \emph{ExactTime}.
With 1--50\,ms random execution, smaller tolerances make cross-index matches unlikely; larger tolerances make them more likely.
\emph{ApproximateTime} simulates developer choices that deviate from strict DAG semantics, as observed in practice.

\textbf{Platform:}
Intel Xeon Silver 4216 (2.10\,GHz, 16C/32T), 124\,GB RAM, Ubuntu 22.04 LTS, Linux 6.8; ROS~2 Humble.

\begin{figure}[tb]
  \centering
  \includegraphics[width=\linewidth]{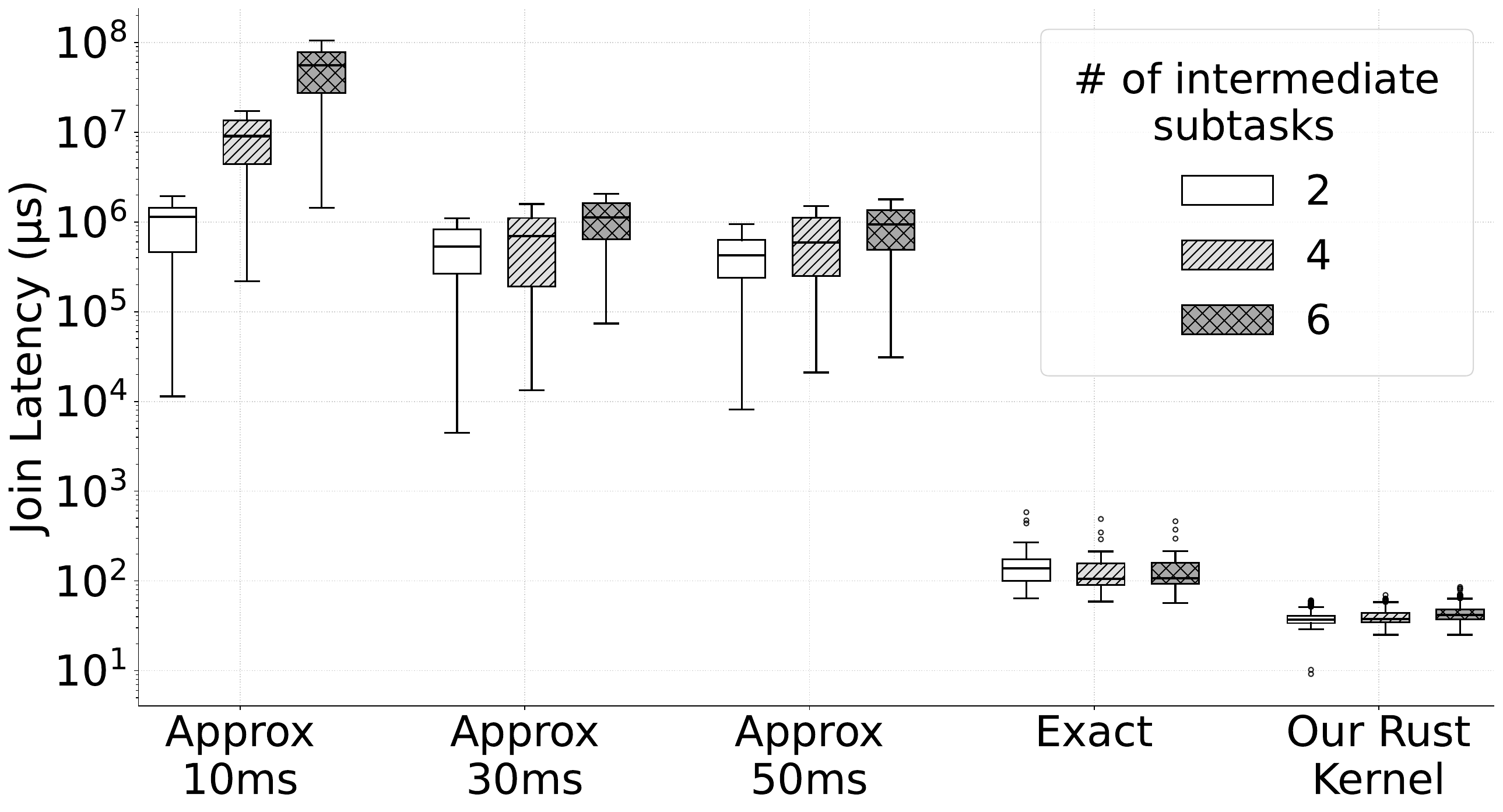}
  \vspace{-6mm}
  \caption{Comparison of join latency.}
  \label{fig-eval}
  \vspace{-5mm}
\end{figure}

\textbf{Results.}
\cref{fig-eval} shows that \emph{ROS~2+ExactTime} and \emph{Case1 kernel} both keep join latency near-minimal; the latter does so by API-enforced semantics.
In contrast, \emph{ROS~2+ApproximateTime} exhibits substantially higher latency, which (i) grows with $N$ (matching becomes harder) and (ii) decreases as \emph{Max Interval Duration} increases (wider tolerance eases matching).
Under \emph{ApproximateTime}, messages accumulate at the join; the resulting wait manifests as higher latency.

\vspace{-1mm}
\section{Conclusion and Future Work} \label{section-conclusion}
\vspace{-1.5mm}
Replacing publish/subscribe with the \TOOLNAME{} (\SHORTNAME{}) API as the middleware interface for component-oriented real-time systems enables schedulers to enforce DAG semantics at the API rather than through programmer discipline.
This alignment with the DAG task model opens a direct path from decades of real-time scheduling theory to practice.

As future work, we are implementing a \SHORTNAME{}-based DAG scheduler on Linux \textit{sched\_ext} (\cref{fig-os-design} right).
Second, we will extend \SHORTNAME{} to cover scheduling factors, including synchronized shared-variable accesses and blocking system calls.
In particular, common ROS~2 idioms where callbacks within the same node pass data via shared variables will be made explicit by \SHORTNAME{}, making such flows first-class citizens.

\vspace{-1mm}
\section{Acknowledgment}
\vspace{-1mm}
This research is based on results obtained from a project, JPNP21027, subsidized by the New Energy and Industrial Technology Development Organization.
Green Innovation Fund Projects / Development of In-vehicle Computing and Simulation Technology for Energy Saving in Electric Vehicles.

\begin{comment}
\section*{APPENDIX}
Appendixes should appear before the acknowledgment.
\end{comment}

\begin{comment}
\section*{ACKNOWLEDGMENT}
\end{comment}

\begin{comment}
\IEEEtriggeratref{1}
\IEEEtriggercmd{\vspace{-2mm}}
\end{comment}
\bibliographystyle{IEEEtran} % IEEEスタイルの参考文献フォーマット
\bibliography{references}     % references.bibファイルを参照

\end{document}